\documentclass[iop,apj]{emulateapj}

\usepackage{apjfonts}
\renewcommand{\prl}{Phys. Rev. Lett.}
\renewcommand{\apjs}{Astrophys. J. Supp.}
\renewcommand{\apjl}{Astrophys. J.}
\renewcommand{\apj}{Astrophys. J.}
\renewcommand{\jgr}{J. Geophys. R.}
\renewcommand{\grl}{Geophys. R. Lett.}
\renewcommand{\mnras}{Mon. Not. R. Astron. Soc.}
\renewcommand{\apjl}{Astrophys. J.}
\providecommand{\ang}{Ann. Geophys.}
\providecommand{\nat}{Nature}
\providecommand{\aass}{Astron. Astrophys. Suppl. Ser.}
\providecommand{\lrsp}{Living Rev. Solar Phys.}
\providecommand{\araa}{Ann. Rev. Astron. Astrophys.}
\providecommand{\ppcf}{Plasma Phys. Control. Fusion}
\providecommand{\ssr}{Space Sci. Rev.}
\providecommand{\aipcp}{AIP Conf. Proc.}
\providecommand{\ansd}{Dok. Akad. Nauk SSSR}
\providecommand{\pof}{Phys. Fluids}
\providecommand{\npg}{Nonlin. Proc. Geophys.}
\providecommand{\jopp}{J. Plasma Phys.}
\providecommand{\rslpsa}{Proc. R. Soc. A}

\shorttitle{3D STRUCTURE OF SOLAR WIND TURBULENCE}
\shortauthors{CHEN ET AL.}
\slugcomment{}

\begin{document}
\title{Three-Dimensional Structure of Solar Wind Turbulence}
\author{C.~H.~K.~Chen\altaffilmark{1}, A.~Mallet\altaffilmark{2}, A.~A.~Schekochihin\altaffilmark{2}, T.~S.~Horbury\altaffilmark{3}, R.~T.~Wicks\altaffilmark{4}, S.~D.~Bale\altaffilmark{1,5}}
\affil{$^1$Space Sciences Laboratory, University of California, Berkeley, California 94720, USA; chen@ssl.berkeley.edu}
\affil{$^2$Rudolf Peierls Centre for Theoretical Physics, University of Oxford, Oxford OX1 3NP, UK}
\affil{$^3$The Blackett Laboratory, Imperial College London, London SW7 2AZ, UK}
\affil{$^4$Heliophysics Science Division, NASA Goddard Space Flight Center, Greenbelt, Maryland 20771, USA}
\affil{$^5$Physics Department, University of California, Berkeley, California 94720, USA}
\begin{abstract}
We present a measurement of the scale-dependent, three-dimensional structure of the magnetic field fluctuations in inertial range solar wind turbulence with respect to a local, physically motivated coordinate system. The Alfv\'enic fluctuations are three-dimensionally anisotropic, with the sense of this anisotropy varying from large to small scales. At the outer scale, the magnetic field correlations are longest in the local fluctuation direction, consistent with Alfv\'en waves. At the proton gyroscale, they are longest along the local mean field direction and shortest in the direction perpendicular to the local mean field and the local field fluctuation. The compressive fluctuations are highly elongated along the local mean field direction, although axially symmetric perpendicular to it. Their large anisotropy may explain why they are not heavily damped in the solar wind.
\end{abstract}

\section{Introduction}

The solar wind is a weakly collisional plasma \citep[e.g.,][]{kasper08} that is ubiquitously observed to be in a turbulent state \citep{tu95,goldstein95a,horbury05,bruno05a,petrosyan10,matthaeus11}. Much progress has been made in understanding the nature of this turbulence since the first direct spacecraft observations \citep[e.g.,][]{siscoe68,coleman68} but many aspects remain to be fully understood. In particular, the three-dimensional (3D) structure has been poorly characterized. Here, we use a new single-spacecraft technique to measure the 3D structure of turbulence in the fast solar wind.

Turbulence is usually modeled as a local cascade of fluctuations from large to small scales, forming an inertial range. In the solar wind, most of the energy at large scales is in Alfv\'enic fluctuations \citep{belcher71,bruno85,horbury95,bale05}, which have magnetic field and velocity fluctuations perpendicular to the magnetic field direction \citep{alfven42}. Early isotropic magnetohydrodynamic (MHD) turbulence theories \citep{iroshnikov63,kraichnan65} based on Kolmogorov scaling arguments \citep{kolmogorov41a} predict that the energy spectrum of weak Alfv\'enic turbulence is $E(k)\sim k^{-3/2}$, where $k$ is the wavenumber of the fluctuations. Although 1D velocity power spectra in the solar wind at 1 AU display this scaling \citep{mangeney01,podesta07a,salem09,chen11b,boldyrev11}, the magnetic field has a $k^{-5/3}$ scaling \citep[e.g.,][]{matthaeus82a,smith06a,chen11b,boldyrev11}.

It was later realized \citep{montgomery81,shebalin83} that the magnetic field direction can induce anisotropy in plasma turbulence. It was then proposed \citep{higdon84,goldreich95} that Alfv\'enic turbulence tends towards a state of critical balance, in which the timescale of the Alfv\'enic fluctuations propagating along the magnetic field is equal to the timescale of their nonlinear decay. This produces a spectrum perpendicular to the local magnetic field of $E(k_\perp)\sim k_\perp^{-5/3}$, a parallel spectrum of $E(k_\parallel)\sim k_\parallel^{-2}$ and local wavevector scaling $k_\parallel\sim k_\perp^{2/3}$. Solar wind turbulence measurements show evidence for both wavevector anisotropy of the form $k_\perp>k_\parallel$ \citep{crooker82,bieber96,leamon98a,horbury08,podesta09a,wicks10a,chen11a,wicks11a,horbury11} and a steeper spectral index parallel to the local magnetic field \citep{horbury08,podesta09a,luo10,wicks10a,chen11a,wicks11a,horbury11}.

\begin{figure}
\noindent\includegraphics[width=\columnwidth]{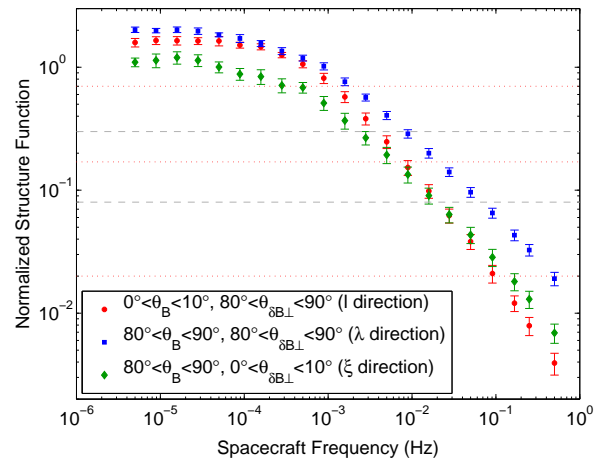}
\noindent\caption{\label{f1}Normalized $\mathbf{B}$-trace structure function in three orthogonal directions. The grey dashed lines indicate the range of values over which spectral indices were fitted. The red dotted lines correspond to the eddy shapes in Fig.~\ref{f4}.}
\end{figure}

Critical balance theory was later extended to allow for the possibility that Alfv\'enic turbulence is 3D anisotropic \citep{boldyrev06}. The two special orthogonal directions are the mean magnetic field $\mathbf{B}_0$ and the perpendicular magnetic field fluctuation $\delta\mathbf{B}_\perp$. The theory assumes that the magnetic field and velocity fluctuations align to within a scale dependent angle $\theta_{vb}$, which makes them 3D anisotropic: $l>\xi>\lambda$, where $l$, $\xi$ and $\lambda$ are their correlation lengths in the mean field direction $\mathbf{B}_0$, in the $\delta\mathbf{B}_\perp$ direction and perpendicular to both, respectively. The local spectra implied by the theory in these three directions are $E(k_l)\sim k_l^{-2}$, $E(k_\xi)\sim k_\xi^{-5/3}$ and $E(k_\lambda)\sim k_\lambda^{-3/2}$. The $\xi$-direction scaling follows from substituting $\xi\sim\lambda^{3/4}$ into $\delta v\sim\lambda^{1/4}$ from \citet{boldyrev06}, giving $\delta v\sim\xi^{1/3}$, corresponding to a local $-5/3$ spectrum.

Scale dependent alignment has been reported in the solar wind at large scales but is difficult to measure deep in the inertial range due to instrumental limitations \citep{podesta09e}. A recent multi-spacecraft measurement of the turbulent energy distribution in the near-Earth solar wind suggested that there was anisotropy with respect to global directions of the system, such as the global mean field, solar wind flow or the bow shock \citep{narita10a,narita10b}. As far as we are aware, there has not yet been a measurement of the 3D structure of solar wind turbulence in a local, scale-dependent coordinate system $(l,\xi,\lambda)$.

Although inertial range solar wind turbulence is predominantly Alfv\'enic, there is also a non-negligible spectrum of compressive fluctuations $\delta B_\parallel$ and $\delta n$, where $n$ is the number density \citep[e.g.,][]{marsch90b,tu94,bavassano04,hnat05,kellogg05,issautier10,chen11b,chen12a}. The nature of these fluctuations is debated \citep{matthaeus91,lithwick01,kellogg05,schekochihin09,howes12,klein12}, in particular, the reason why they are not heavily damped. Their structure has been less comprehensively characterized than the Alfv\'enic turbulence, although measurements in the magnetosheath show that there is some degree of 2D anisotropy \citep{alexandrova08c,he11b}.

In this paper, we present measurements of the scale-dependent 3D structure of the Alfv\'enic and compressive magnetic field fluctuations with respect to a new local coordinate system and discuss the implications for our understanding of plasma turbulence.

\begin{figure}
\noindent\includegraphics[width=\columnwidth]{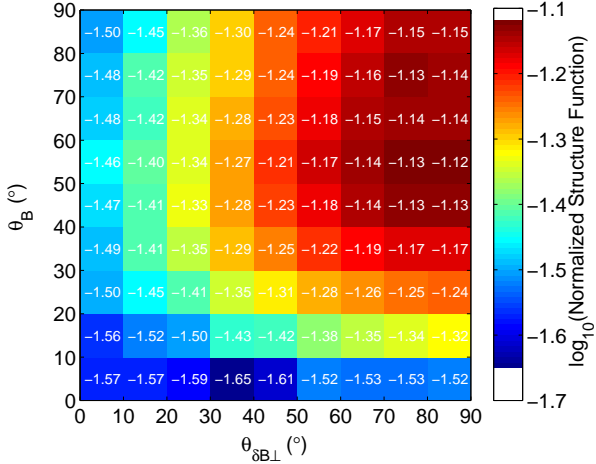}
\noindent\caption{\label{f2}Normalized $\mathbf{B}$-trace structure function at $1.5\times10^{-2}$ Hz as a function of $\theta_B$ and $\theta_{\delta B_\perp}$.}
\end{figure}

\section{Method}

In the analysis, fast solar wind data from the Ulysses spacecraft \citep{wenzel92} during a polar pass between 1.4 and 2.6 AU in days 100--299 of 1995 was used. The magnetic field data from VHM \citep{balogh92} was at 1 sec resolution and the velocity data from SWOOPS \citep{bame92} was at 4 min resolution. The average solar wind speed was $\approx$ 780 km s$^{-1}$ and the outer scale cross-helicity was moderately high, $\sigma_{\mathrm{c}}\approx$ 0.6 (other plasma parameters for this stream are given in \citet{wicks10a}). The data was split into 10 equal length intervals for the analysis.

For each 20 day interval, 21 logarithmically spaced spacecraft-frame frequencies at which to measure the power levels, over the range $5\times 10^{-6}$ Hz $\leq f_{\mathrm{sc}}\leq 5\times 10^{-1}$ Hz, were chosen. For each of these frequencies, the pairs of magnetic field measurements, $\mathbf{B}_1$ and $\mathbf{B}_2$, with the time lag $1/f_{\mathrm{sc}}$ were selected. For each pair, the contribution to the second order $\mathbf{B}$-trace structure function $\sum\limits_i\left(B_{1,i}-B_{2,i}\right)^2$, where $i$ is the component of the magnetic field, and the contribution to the second order $|\mathbf{B}|$ structure function $\left(|\mathbf{B}_1|-|\mathbf{B}_2|\right)^2$ were calculated. Since most of the energy is in the perpendicular fluctuations \citep[e.g.,][]{belcher71}, the $\mathbf{B}$-trace spectrum is a good proxy for the Alfv\'enic $\delta\mathbf{B}_\perp$ spectrum and since $|\mathbf{B}|=|\mathbf{B}_0+\delta\mathbf{B}|\approx\sqrt{|\mathbf{B}_0|^2+2\mathbf{B}_0\cdot\delta\mathbf{B}}\approx|\mathbf{B}_0|+\delta B_{\parallel}$, the $|\mathbf{B}|$ spectrum is a good proxy for the compressive $\delta B_\parallel$ spectrum in the inertial range, where $|\delta\mathbf{B}|<|\mathbf{B}_0|$.

It has been shown that using a local scale-dependent 2D coordinate system is important for testing theoretical predictions of spectral anisotropy \citep{cho00,maron01,horbury08,chen10b,chen11a}. Here, this is extended by defining a local scale-dependent 3D coordinate system. For each pair of points, the local mean field $\mathbf{B}_{\mathrm{local}}=\left(\mathbf{B}_1+\mathbf{B}_2\right)/2$ and the local perpendicular fluctuation direction $\mathbf{B}_{\mathrm{local}}\times\left[\left(\mathbf{B}_1-\mathbf{B}_2\right)\times\mathbf{B}_{\mathrm{local}}\right]$ were calculated. The angle between $\mathbf{B}_{\mathrm{local}}$ and the mean solar wind velocity (which is the sampling direction), $\theta_B$, and the angle between the local perpendicular fluctuation and the component of the solar wind velocity perpendicular to $\mathbf{B}_{\mathrm{local}}$, $\theta_{\delta B_\perp}$, were then found.

\begin{figure}
\noindent\includegraphics[width=\columnwidth]{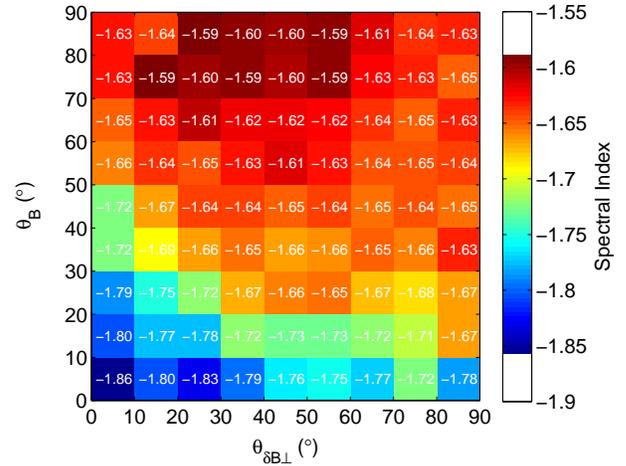}
\noindent\caption{\label{f3}$\mathbf{B}$-trace spectral index between normalized structure function values of 0.08 and 0.3 as a function of $\theta_B$ and $\theta_{\delta B_\perp}$.}
\end{figure}

\begin{figure*}
\noindent\includegraphics[scale=0.5]{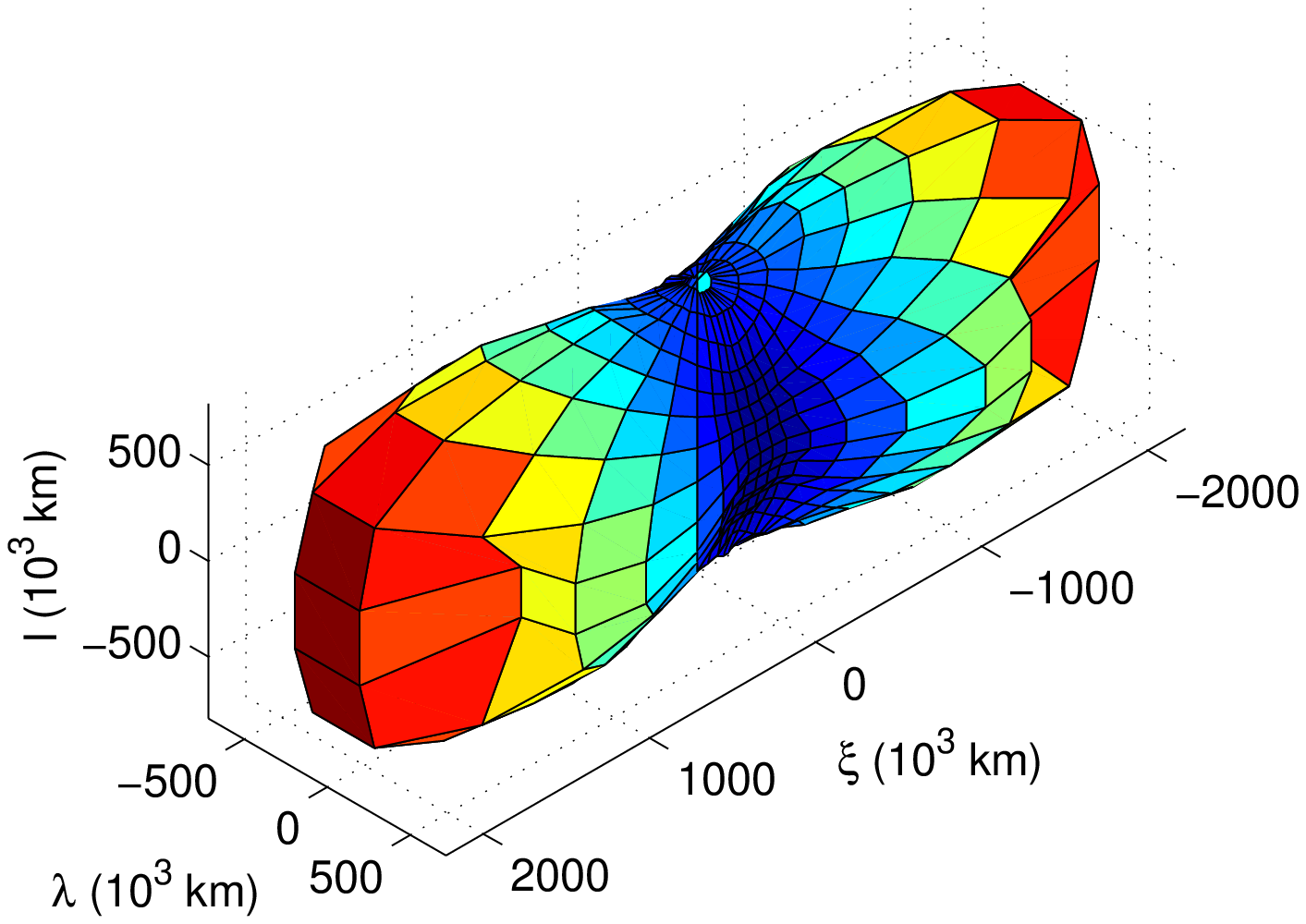}
\noindent\includegraphics[scale=0.5]{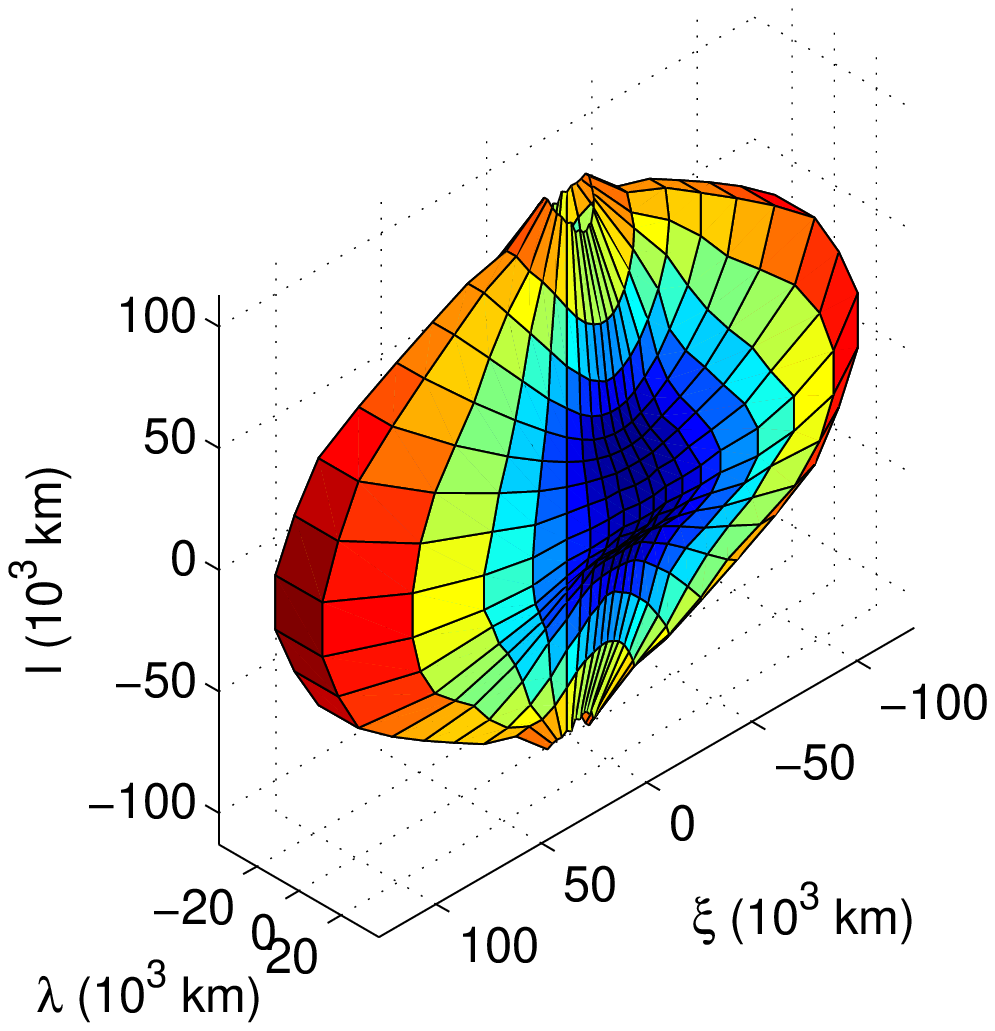}
\noindent\includegraphics[scale=0.5]{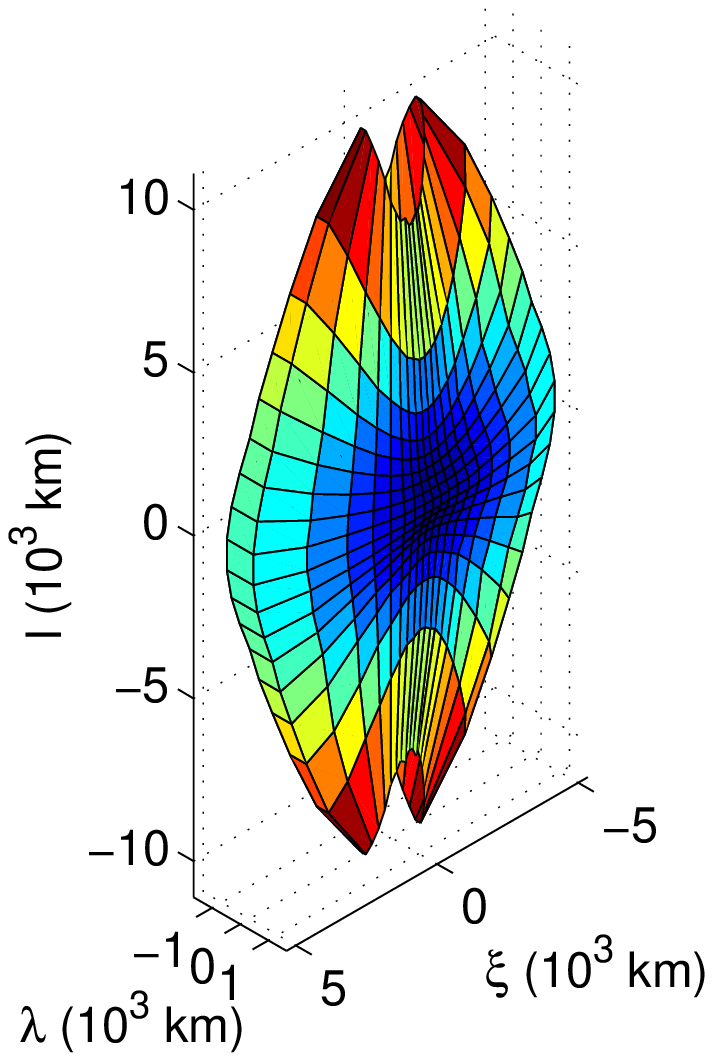}
\noindent\caption{\label{f4}Surfaces of constant $\mathbf{B}$-trace power (statistical Alfv\'enic eddy shapes) from large (left) to small (right) scales, in which color represents distance from the origin. The normalized power levels are 0.7, 0.17 and 0.02 as marked with red dotted lines on Fig.~\ref{f1}. The typical proton gyroradius is $\approx$ 360 km.}
\end{figure*}

An orthogonal spherical polar coordinate system was defined, in which $f_{\mathrm{sc}}$ is the radial coordinate, $\theta_B$ is the polar angle and $\theta_{\delta B_\perp}$ is the azimuthal angle. It is in this local coordinate system that the 3D anisotropy of the turbulence was measured. The structure function contributions for each $f_{\mathrm{sc}}$ were binned in 10$^\circ$ bins of $\theta_B$ and $\theta_{\delta B_\perp}$ and the mean value in each bin was calculated. While the structure functions conditioned to the local field direction in this way may not be purely second order \citep{matthaeus12}, they are thought to be the relevant quantities in critical balance theories \citep{cho00,horbury08,chen11a}. Any angles greater than 90$^\circ$ were reflected below 90$^\circ$ to improve accuracy for scaling measurements. Reflection in $\theta_{\delta B_\perp}$ was found to be a good approximation; while there were few points to check the validity of reflection in $\theta_B$, the assumption seems reasonable \citep{podesta09a}.

Taylor's hypothesis \citep{taylor38} can be assumed for this analysis: since the speed of the solar wind moving past the spacecraft is more than 10 times the Alfv\'en speed in this interval \citep{wicks10a}, temporal variations measured by the spacecraft, $1/f_{\mathrm{sc}}$, correspond to spatial variations in the plasma $v_{\mathrm{sw}}/f_{\mathrm{sc}}$, where $v_{\mathrm{sw}}$ is the solar wind speed. This has been shown to be a good approximation \citep{narita10d}.

\section{Results}
\subsection{Alfv\'enic Fluctuations}

Fig.~\ref{f1} shows the $\mathbf{B}$-trace structure function (``power'') as a function of spaceraft-frame frequency for three angle bins corresponding to the $\mathbf{B}_{\mathrm{local}}$ direction (red circles), the $\delta\mathbf{B}_\perp$ direction (green diamonds) and the direction perpendicular to both (blue squares). Each value is the mean calculated from the 10 intervals and the error bars are $2\sigma$, where $\sigma$ is the standard error of the mean. Before averaging, the structure functions of each interval were normalized to the square of the mean field strength over the interval $\left<|\mathbf{B}|\right>^2$ to account for the varying power levels due to the spacecraft orbit. The typical proton gyroscale corresponds to a spacecraft-frame frequency $\approx$ 0.3 Hz.

The perpendicular (blue) curve is characteristic of fast solar wind: shallow in the low frequency $1/f_{\mathrm{sc}}$ range \citep{matthaeus86} and steeper in the higher frequency inertial range. The parallel (red) curve also matches previous parallel spectrum measurements, following the perpendicular curve at low frequencies, then becoming steeper than it in the inertial range \citep{wicks10a}. The $\delta\mathbf{B}_{\perp}$ (green) curve has not previously been measured and describes how the 3D anisotropy evolves in the turbulent cascade. At large scales it has a smaller value than the other structure functions, which is consistent with this range consisting of Alfv\'en waves \citep{belcher71}, since they have wavevectors in the plane perpendicular to $\delta\mathbf{B}_\perp$. It also remains smaller than the perpendicular structure function throughout the cascade but becomes larger than the parallel one at $\approx 3\times 10^{-2}$ Hz.

For each 20 day interval, a power law was fitted to the normalized structure functions between values of 0.08 and 0.3 (marked as grey dashed lines) in each angle bin. A fixed power range, rather than a fixed $f_{\mathrm{sc}}$ range, was used so that the scaling was measured for the same set of fluctuations \citep{chen10a}. For each angle bin, the fit to the structure function was evaluated at $1.5\times10^{-2}$ Hz to give the 3D power anisotropy and the mean of the 10 intervals is shown in Fig.~\ref{f2}. The typical standard error of the log of the mean is between 0.05 and 0.07. It can be seen that the power increases with both $\theta_B$ and $\theta_{\delta B_\perp}$, indicating 3D anisotropy, and seems to peak near $\theta_B=60^\circ$, $\theta_{\delta B_\perp}=90^\circ$.

Each fitted power law index was converted to a spectral index by subtracting 1 \citep{monin75} and the 3D spectral index anisotropy is shown in Fig.~\ref{f3}. The typical standard error of the mean is 0.01 or 0.02, although the actual uncertainty may be larger due to systematic effects, such as the finite frequency response of the structure functions. The steepening towards small $\theta_B$ \citep{horbury08} can be seen but there appears to be little variation with $\theta_{\delta B_\perp}$ at large $\theta_B$.

To visualize how the 3D anisotropy varies through the turbulent cascade, surfaces of constant power were calculated. At a selected structure function value, the corresponding frequency in each angle bin was found through linear interpolation and the scales corresponding to these frequencies were calculated using Taylor's hypothesis. The scales, together with the angles $\theta_B$ and $\theta_{\delta B_\perp}$, were converted into Cartesian coordinates $(l,\xi,\lambda)$ and the surfaces of constant power (at structure function values marked by red dotted lines in Fig.~\ref{f1}) are shown in Fig.~\ref{f4}. They have been reflected into the other seven octants under the assumption of reflectional symmetry (see earlier). These statistical surfaces can loosely be considered as average eddy shapes (although they are not eddies in the dynamical sense). It can be seen that they change from being extended in the $\delta\mathbf{B}_\perp$ direction in the large scale Alfv\'en wave range ($\xi>l,\lambda$) to being 3D anisotropic close to the proton gyroscale ($l>\xi>\lambda$).

\begin{figure}
\noindent\includegraphics[width=\columnwidth]{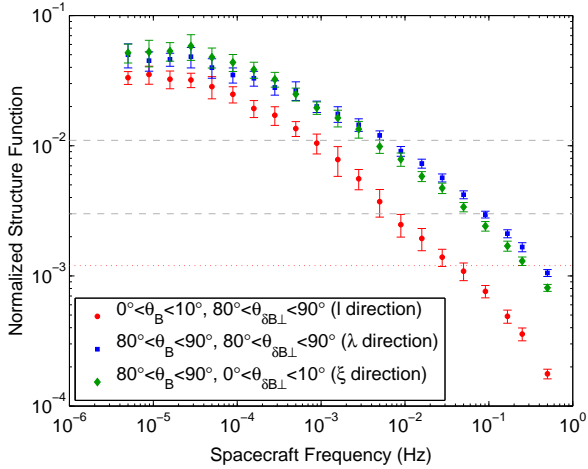} 
\noindent\caption{\label{f5}Normalized $|\mathbf{B}|$ structure function in three orthogonal directions. The grey dashed lines show the range of values over which spectral indices were fitted. The red dotted line corresponds to the eddy shape in Fig.~\ref{f6}.}
\end{figure}

\subsection{Compressive Fluctuations}

\begin{figure}
\noindent\includegraphics[width=\columnwidth]{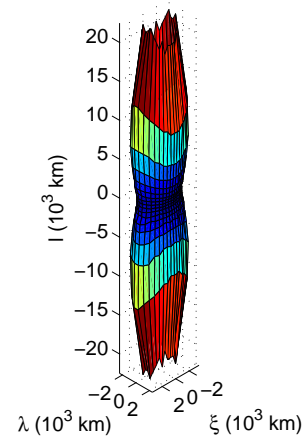}
\noindent\caption{\label{f6}Surface of constant $|\mathbf{B}|$ power (statistical compressive eddy shape) at small scales. The normalized power level is $1.2\times10^{-3}$ as marked with a red dotted line on Fig.~\ref{f5}.}
\end{figure}

The results of a similar analysis for $|\mathbf{B}|$ are shown in Fig.~\ref{f5} (where the structure functions have been normalized in the same way as for the Alfv\'enic fluctuations in Fig.~\ref{f1}) and Fig.~\ref{f6}, which is the surface of constant normalized power of $1.2\times10^{-3}$ (marked as a red dotted line in Fig.~\ref{f5}). It can be seen that the structure of the compressive fluctuations is different to the Alfv\'enic fluctuations: there is no anisotropy in the plane perpendicular to the mean field, meaning that the compressive fluctuations do not depend on the polarization of the Alfv\'enic fluctuations. Also, they are more elongated along the mean field direction than the Alfv\'enic fluctuations: for a given perpendicular scale $\lambda$, the ratio $l/\lambda$ is at least 2 or 3 times larger. Due to limited angular resolution this is a lower limit; by extrapolating the shape in Fig.~\ref{f6} one could imagine that they are even more extended than can currently be measured.

The spectral indices of $|\mathbf{B}|$ for normalized powers between $3\times10^{-3}$ and $1.1\times10^{-2}$ are between --1.58 and --1.42 in all angle bins, with a typical standard error of the mean of 0.02. This is different to the slow solar wind, where spectral indicies close to --5/3 are observed \citep{chen11b}. This difference has also been noticed in the electron density spectrum \citep{issautier10}, although the reason is not well understood. If the compressive fluctuations are indeed very anisotropic, then we would not expect to measure the true parallel spectral index with the current angular resolution, which may explain the presence of anisotropic structures yet no significant anisotropic scaling.

\section{Discussion}

We have shown that the Alfv\'enic turbulence is locally anisotropic in the plane perpendicular to the mean field. Since the direction of the anisotropy is associated with $\delta\mathbf{B}_\perp$, the question naturally arises to what extent this anisotropy is a reflection of the solenoidality of the magnetic field \citep{turner11}. While the magnetic field has zero divergence at each point, this does not imply that the correlation length along $\delta\mathbf{B}_\perp$ is infinite, because at any given scale we are considering finite field increments, not derivatives.

The effect of solenoidality can be derived from knowledge of the probability density function (PDF) of $\delta\mathbf{B}_\perp$. Since we are considering the conditional structure function at each scale $\langle \delta B_\perp^2|\theta_{\delta B_\perp}\rangle$, the joint PDF $p(\delta B_\perp,\theta_{\delta B_\perp})$, or, equivalently, the PDF of the vector $\delta\mathbf{B}_\perp$ is required. As a simple illustration, consider the case where this PDF is Gaussian and, therefore, fully determined by the second-order longitudinal correlation function $C_{LL}(r)=\langle |\delta\mathbf{B}_\perp\cdot\mathbf{r}/r|^2\rangle$, where $\mathbf{r}$ is the point separation in the perpendicular plane. The conditional structure function becomes
\begin{equation}
\langle \delta B_\perp^2(r)|\theta_{\delta B_\perp}\rangle = \frac{2C_{LL}(r)C_{TT}(r)}{C_{LL}(r)\sin^2\theta_{\delta B_\perp} + C_{TT}(r)\cos^2\theta_{\delta B_\perp}},
\end{equation}
where the transverse correlation function is $C_{TT}(r) = [r C_{LL}(r)]'$ from solenoidality \citep{batchelor53}. If, in the inertial range, $C_{LL}(r)\propto r^\alpha$, then
\begin{equation}
\langle \delta B_\perp^2(r)|\theta_{\delta B_\perp}\rangle \propto \frac{r^\alpha}{1+\alpha \cos^2\theta_{\delta B_\perp}}.
\end{equation}
Therefore, the ratio of the correlation scales along and across the fluctuation direction for a given structure function value is scale-independent and equal to $\xi/\lambda = (1+\alpha)^{1/\alpha}$, which, for the measured value of $\alpha\approx 2/3$, gives an anisotropy in the perpendicular plane of $\approx 2.15$. Since the measured anisotropy is larger than this (varying between 3.2 and 3.8), non-Gaussianity is required to explain the observations. Any scale-dependent alignment, e.g., the dynamical alignment of \citet{boldyrev06}, is likely to require non-Gaussianity and therefore be closely related to the intermittency of the turbulence.

Our results show some important differences to an earlier study, which suggested that the small scale fluctuations are longest-correlated in one of the perpendicular directions and that the spectral index is different in all three directions \citep{narita10a}. Possible reasons for this include the different coordinate system used (global rather than local), the presence of foreshock effects in \citet{narita10a} or the different assumptions in the measurement technique.

The fact that the compressive fluctuations are very elongated is consistent with the prediction, based on gyrokinetic theory, that they are passive to the Alfv\'enic fluctuations, but have no parallel cascade along the exact magnetic field lines \citep{schekochihin09}. This may explain why there is a compressive cascade in the solar wind: the compressive fluctuations are expected to be damped at a rate proportional to their parallel wavenumber $\gamma\sim k_\parallel$ \citep{barnes66,schekochihin09,klein12} but if $k_\parallel$ is very small then they are not heavily damped and can cascade nonlinearly. An alternative explanation is that the less anisotropic compressive fluctuations are generated but are quickly damped, leaving the highly elongated structures to be observed.

\acknowledgments
This work was supported by NASA contract NNN06AA01C, NASA grant NNX09AE41G and the Leverhulme Trust Network for Magnetized Plasma Turbulence. Ulysses data was obtained from CDAWeb ({http://cdaweb.gsfc.nasa.gov}).

\end{document}